# A Spin-dependent Machine Learning Framework for Transition Metal Oxide Battery Cathode Materials

Taiping Hu[1,2], Teng Yang[3], Jianchuan Liu[4], Bin Deng[5], Zhengtao Huang[3], Xiaoxu Wang[5], Fuzhi Dai[2], Guobing Zhou[6], Fangjia Fu[2,7], Ping Tuo[2], Ben Xu[*,3], and Shenzhen Xu[*,1,2]

[1]Beijing Key Laboratory of Theory and Technology for Advanced Battery Materials, School of Materials Science and Engineering, Peking University, Beijing 100871, People's Republic of China

[2]AI for Science Institute, Beijing 100084, People's Republic of China

[3]Graduate School of China Academy of Engineering Physics, Beijing 100088, People's Republic of China

[4] HEDPS, CAPT, College of Engineering and School of Physics, Peking University, Beijing 100871, People's Republic of China

[5]DP Technology, Beijing 100080, People's Republic of China

[6] Institute of Advanced Materials, Jiangxi Normal University, Nanchang 330022, People's Republic of China

[7] School of Mathematical Sciences, Peking University, Beijing 100871, People's Republic of China

Corresponding author: bxu@gscaep.ac.cn, xushenzhen@pku.edu.cn,

# Abstract

Owing to the trade-off between the accuracy and efficiency, machine-learning-potentials (MLPs) have been widely applied in the battery materials science, enabling atomic-level dynamics description for various critical processes. However, the challenge arises when dealing with complex transition metal (TM) oxide cathode materials, as multiple possibilities of *d*-orbital electrons localization often lead to convergence to different spin states (or equivalently local minimums with respect to the spin configurations) after *ab initio* self-consistent-field calculations, which causes a significant obstacle for training MLPs of cathode materials. In this work, we introduce a solution by incorporating an additional feature - atomic spins - into the descriptor, based on the pristine deep potential (DP) model, to address the above issue by distinguishing different spin states of TM ions. We demonstrate that our proposed scheme provides accurate descriptions for the potential energies of a variety of representative cathode materials, including the traditional $Li_xTMO_2$ (TM=Ni, Co, Mn, $x$=0.5 and 1.0), Li-Ni anti-sites in $Li_xNiO_2$ ($x$=0.5 and 1.0), cobalt-free high-nickel $Li_xNi_{1.5}Mn_{0.5}O_4$ ($x$=1.5 and 0.5), and even a ternary cathode material $Li_xNi_{1/3}Co_{1/3}Mn_{1/3}O_2$ ($x$=1.0 and 0.67). We highlight that our approach allows the utilization of all *ab initio* results as a training dataset, regardless of the system being in a spin ground state or not. Overall, our proposed approach paves the way for efficiently training MLPs for complex TM oxide cathode materials.

# Introduction

Lithium-ion batteries (LIBs) have revolutionized portable electronic devices and electric vehicles due to their high energy density, light weight, and long cycle life.[1-4] The cathode, as one of the fundamental components of LIBs, plays a crucial role in determining battery's performances and costs.[5, 6] However, improving the structural stability under high operating voltage for almost all cathode materials remains challenging.[7-10] Experimentally, many strategies, often relying on trial and error, have been devoted to enhance the structural stability.[8, 10-13] Benefiting from computational capacities' improvement, atomic-level simulations become increasingly valuable in helping interpret experimental observations and guide materials' designs.[14-16] Molecular dynamics (MD) simulations, in particular, provide comprehensive dynamic evolutions at the atomic scale, making them widely used in LIB researches.[17-19]

Owing to the trade-off between the accuracy and efficiency, machine learning potentials (MLPs)[20-22] have been found extensive applications recently in describing the complete dynamics of critical processes in LIBs. For example, MLPs have been successfully applied to the solid electrolyte interface[23], the Si and Li metal anodes[24, 25], and organic/solid-state electrolytes[26, 27]. However, to the best of our knowledge, only a few studies reported MLPs for Li battery cathodes,[28, 29] primarily because that *ab initio* calculations suffer numerous challenges when dealing with complicated transition metal (TM) oxides cathodes. On the one hand, TM ions undergo changes in their oxidation states during the lithium insertion/extraction process. There are multiple possible magnetic configurations (e.g., high-spin (HS), intermediate-spin (IS), and low-spin (LS) states) even for a TM ion at a specific valence state. On the other hand, the density functional theory with the Hubbard $U^{30}$ (DFT+U) method is usually applied to those strongly correlated systems to correct the so-called self-interaction errors. A relatively random initial wavefunction or small structural perturbations could result in different localization of *d*-orbital electrons on the TM ion sites/orbitals, and converging to different potential energy surfaces (with respect to the degrees of freedom of the TM ions' spin states). The above issues pose significant challenges for constructing cathodes' MLPs. Therefore, it is highly desired to seek a systematic approach to effectively address those challenges.

In our previous work, we developed a workflow to identify the magnetic ground state for the LiCoO$_2$ cathode.[28] Subsequently, we successfully constructed the deep potential (DP) model[31], one of the popular MLPs, for the Li$_x$CoO$_2$ material with different phases and a wide range of concentrations. Specifically, we at the beginning confirmed the magnetic ground state and then excluded the non-ground states (with respect to spin configurations) in DFT+U single-point calculations. However, we realized that the above workflow has two drawbacks: (1) it relies on manual interventions, which limits its applications to more complex TM oxide cathode materials. (2) it leads to substantial data wastage due to the exclusion of the systems at non-ground spin states, which may require an expensive cost for constructing ground-state MLPs. Here we emphasize that the DFT+U results converging to non-ground state do not mean that the corresponding electronic-structure calculations are incorrect. As long as the self-consistent-field calculations converge successfully, the obtained electronic-structure results and the associated spin configurations are valid, which are just likely to be local minimums but not global minimums with respect to the degrees of freedom of the TM ions' spin states. A question then naturally arises: how can we automatically determine the magnetic ground state while fully utilizing the data produced by DFT+U single-point calculations in the MLPs training process?

Recently, Xu et al[32] developed the DeePSPIN model to simulate the simultaneous evolution of both the lattice and spin in magnetic materials. Although the DeePSPIN model's primary goal is to handle lattice-spin interactions in complex magnetic systems, its key idea of incorporating the spin into the descriptor inspires us to resolve the cathode materials' MLPs training challenges mentioned above. In fact, the spin configuration of TM ions is an additional feature beyond the atomic coordinates for TM oxide cathodes. In other words, for a specific structure, multiple spin states may exist, corresponding to the spin ground state and "excited" states (illustrated in **Figure 1a**). Therefore, in this work, by employing a collinear DeePSPIN model's framework, we also integrate the spin into the descriptor (as depicted in **Figure 1b**), aiming to distinguish different spin states of TM ions for a specific atomic structure of the studied system. We demonstrate that our proposed scheme can provide accurate descriptions for the potential energies of various representative cathode materials, from simple LiTMO$_2$ (TM=Ni, Co, Mn) to complex ternary NCM cathodes. More importantly, the current workflow does not require any manual intervention and can fully utilize all data generated from DFT+U single-point calculations, which avoids identification of ground states and data wastage, thus enabling a more efficient and automated workflow.

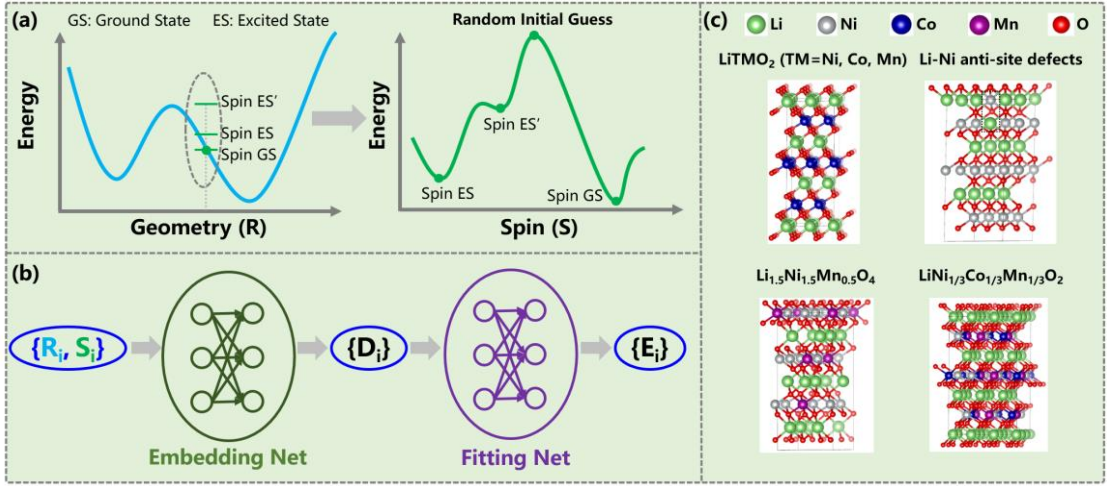

**Figure 1.** (a) Schematic plot of the potential energy surface containing both the degrees of freedom of geometric configurations and spin states. (b) General framework of the DeePSPIN model adopted in this work. (c) Atomic structures of the test systems in this study.

## Methodology

### *The Collinear DeePSPIN Model*

Here we briefly introduce the implementation of the DeePSPIN model.[32] In this framework, a virtual atom $\mathbf{R}'_i$ is introduced near a magnetic atom $\mathbf{R}_i$. The position of the virtual atom is given by the following relationship,

$\mathbf{R}'_i = \mathbf{R}_i + \eta \mathbf{S}_i, i = 1, ..., N$     (1)

where $\mathbf{S}_i$ is a three-dimensional vector in the *non-collinear* framework and $N$ denotes the number of magnetic atoms. $\eta$ is a hyperparameter, called "virtual length", which is used to control the Euclidean distance between the virtual atom $\mathbf{R}'_i$ and the real atom $\mathbf{R}_i$.

Different from the format of magnetic moments data required by the DeePSPIN scheme, all DFT+U calculations in the field of battery cathode simulations are typically performed within the *collinear* framework. Therefore, the atomic spin, denoted as $S_i$, is actually a scalar. The positive and negative values indicate the spin-up and spin-down states, respectively. The formula (1) can be rewritten as,

$$z'_i = z_i + \eta S_i, i = 1, \dots, N \quad (2)$$

where $z'_i$ ($z_i$) represents the z-component of the Cartesian coordinate of virtual (real) atoms, and $S_i$ is the magnetic moment of the *i*-th magnetic atom projected on the *z* direction. In addition, we introduce a constant *d* to the formula (2) to avoid the overlap of the real and virtual atoms when spins are close to zero.

$$z'_i = z_i + \eta S_i + d, i = 1, \dots, N \quad (3)$$

By introducing virtual atoms, we effectively represent information of both the atomic geometries and spin states within the extended atomic coordinates, allowing for a more complete description of the system.

### *Density Functional Theory Calculations*

Spin-polarized density functional theory (DFT) calculations in this work were performed within the Vienna Ab initio Simulation Package (VASP, version 5.4.4).[33, 34] We employed the projector augmented wave (PAW)[35] potentials for modeling the nuclei and the frozen-core electrons of all atoms. The valence electron configurations were $2s^22p^4$ for O, $3d^84s^1$ for Co, $1s^22s^1$ for Li, $3p^63d^64s^1$ for Mn, and $3d^94s^1$ for Ni. We applied the Perdew-Burke-Ernzerhof (PBE) functional[36] with the Hubbard U correction[30] (PBE+U) to the transition metal Co, Ni and Mn. The U values for Co, Ni and Mn were set as 5.14 eV, 6.30 eV and 3.9 eV, obtained from previous works.[37, 38] Unless specifically explained, a dense reciprocal-space mesh with 0.25 Å$^{-1}$ and a 520 eV kinetic energy cutoff for the plane wave basis were employed. The self-consistent field electronic-structure calculations were converged within $10^{-5}$ eV for the total energies.

## *Training of MLPs*

The constructions of all datasets used in this study are described in the Supporting Information (SI). We trained the collinear DeePSPIN model with $4\times10^6$ steps using the deepmd-kit software (version 2.2.2).[39, 40] The embedding network has three layers with 25, 50 and 100 nodes and the fitting network is composed of three layers, each of which has 240 nodes. We used the Adam method[41] to minimize the loss function with an exponentially decay learning rate from $1.00\times10^{-3}$ to $3.51\times10^{-8}$. Due to lack of spin forces' labels in collinear DFT+U calculations, we turn off the spin forces' pre-factor in the loss function during the training process.

## **Results and Discussion**

We first investigate the performance of the pristine DP model (the regular DP model without spin features in the descriptor[31]) for simple $LiTMO_2$ (TM = Ni, Co, Mn) cathodes (atomic structures are displayed in **Figure 1c**) by using the datasets generated by our previously developed workflow,[28] where the ground-state results are identified and imported into the neural network training process. We construct the DP models for the $Li_xCoO_2$, $Li_xMnO_2$ and $Li_xNiO_2$ ($x$ = 0.5 and 1.0). We note that the $Co^{3+}/Ni^{3+}$ and $Co^{4+}/Ni^{4+}$ ions are in the low spin states (the magnetic moments are 0/1 $\mu_B$ and 1/0 $\mu_B$, respectively), while both the $Mn^{3+}$ and $Mn^{4+}$ ions are in the high spin states (the magnetic moments are 4 $\mu_B$ and 3 $\mu_B$, respectively), based on the previous well-known knowledge[28, 37, 42, 43]. We evaluate the performances of the traditional DP models by comparing energies and forces predicted by DFT+U calculations and the DP models. The root-mean-square errors (RMSEs) of energies and forces are 3.2 meV/atom and 115 meV/Å for $Li_xCoO_2$, 3.3 meV/atom and 101 meV/ Å for $Li_xMnO_2$, and 2.9 meV/atom and 114 meV/Å for $Li_xNiO_2$, respectively ($x$ = 0.5 and 1.0, see **Figure 2**). Such small errors show that the pristine DP model can accurately describe the potential energy surfaces of magnetic TM oxide materials as long as the electronic structures of all the configurations in the training datasets are in the spin ground states.

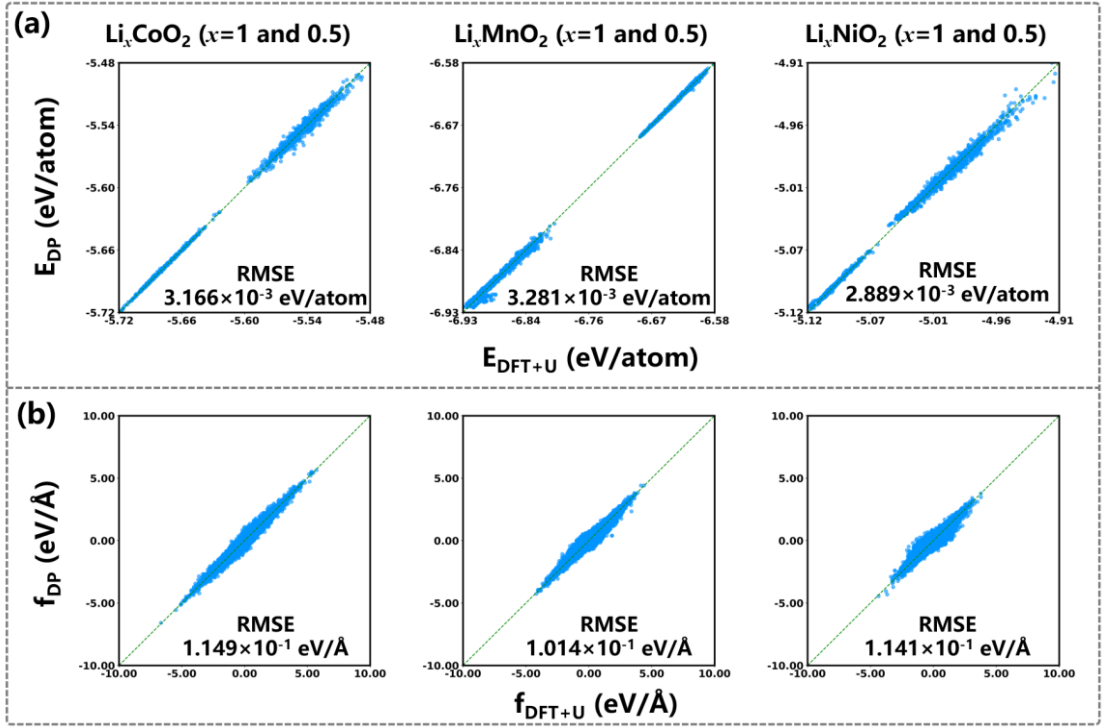

**Figure 2.** Comparisons of the (a) energies and (b) forces predicted by the DFT+U calculations vs. the traditional DP model for $Li_xCoO_2$, $Li_xMnO_2$ and $Li_xNiO_2$ cathodes ($x$ = 0.5 and 1.0), where only the spin ground states' data are taken into consideration.

We obtain the MLPs by fitting a dataset generated by DFT+U single-point calculations and constructing a mapping between atomic coordinates and [energies + forces]. However, training an accurate MLP becomes challenging when the data from multiple potential energy surfaces (labeled by different spin configurations) are mixed together. This situation somehow resembles a one-to-many mapping, where a single atomic configuration corresponds to multiple energetic states which are actually associated with different spin states of the TM ions. To verify the above claim, we add the magnetic excited state data to the existing ground state dataset of the $LiTMO_2$. We consider the high spin states of the $Co^{3+}/Co^{4+}$ (4/5 $\mu_B$) and $Ni^{3+}/Ni^{4+}$ (3/4 $\mu_B$) ions, and the intermediate spin state of the $Mn^{3+}$ (2 $\mu_B$) and the low spin state of the $Mn^{4+}$ ions (1 $\mu_B$). We maintain the same training parameters as before. As expected, we can see that the pristine DP model yields considerably poor predictions for both energies and forces (see **Figure 3**). In particular, the RMSEs of energy comparison are ~ 100 meV/atom for the test systems, which is unacceptable. The MLP's accuracy thus could be significantly affected by the presence of multiple potential energy surfaces associated with different spin states in the training dataset.

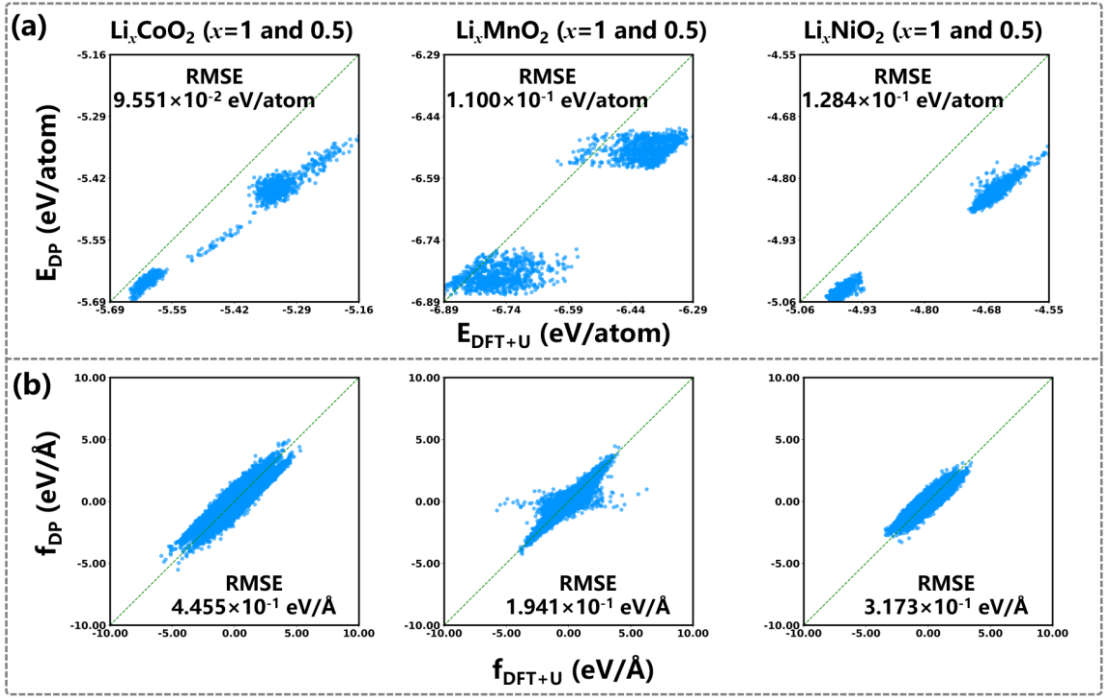

**Figure 3.** Comparisons of the (a) energies and (b) forces predicted by the DFT+U calculations vs. the pristine DP model for $Li_xCoO_2$, $Li_xMnO_2$ and $Li_xNiO_2$ ($x$ = 0.5 and 1.0) cathodes. In addition to the spin ground states' data, the spin excited states' data are also included.

By contrast, in the DeePSPIN model, different spin states of TM ions are represented by virtual atoms with distinct positions. Therefore, even if multiple spin states corresponding to the same geometric structure may be mixed in the training dataset, they could be identified as different input data points. We then retrain the above datasets by using our collinear DeePSPIN model. We can see a remarkable improvement in the predictions for both energies and forces (see **Figure 4**). The RMSEs of energies are ~ 2-3 orders of magnitude lower than those predicted by the pristine DP model. Surprisingly, the energies' RMSEs for $Li_xCoO_2$ and $Li_xNiO_2$, as well as the forces' RMSE for the $Li_xNiO_2$ are even smaller than those given by the DP models that are trained solely on spin ground states' dataset (as shown in **Figure 2**). Such encouraging performance of the collinear DeePSPIN model can be attributed to its ability to distinguish different spin states.

We also realize that the positions of the virtual atoms (representing the information of TM ions' spin states) may affect the accuracy of the collinear DeePSPIN model. We therefore further investigate the dependence of the model's accuracy on two key parameters, $\eta$ and $d$ in the formula (3), involved in constructing virtual atoms. We found that the model's performance remains stable when those two parameters change within a reasonable range (see **Figure S1**). For example, for the $Li_xCoO_2$ case, RMSEs of energies and forces fluctuate in the range of 2.2 – 3.2 meV/atom and 0.15 – 0.19 eV/Å, respectively, when the η and d values continuously increase from 0.1 to 0.5. These results demonstrate that our proposed collinear DeePSPIN model exhibits robustness for training cathode materials' MLPs. Another crucial aspect is that whether the training strategy needs to be modified after introducing an additional degree of freedom in the descriptor. We thus test the impact of the pre-factor of the atomic force in the loss function on the model's accuracy. We find that the force's pre-factor has a negligible effect on the model's performance (see **Table S3**). In other words, we still can employ the training strategy similar with the pristine DP model, which involves progressively increasing the energy's pre-factor and decreasing the force's pre-factor in the loss function, so that the force term dominates initially while the energy becomes more important at end.[31]

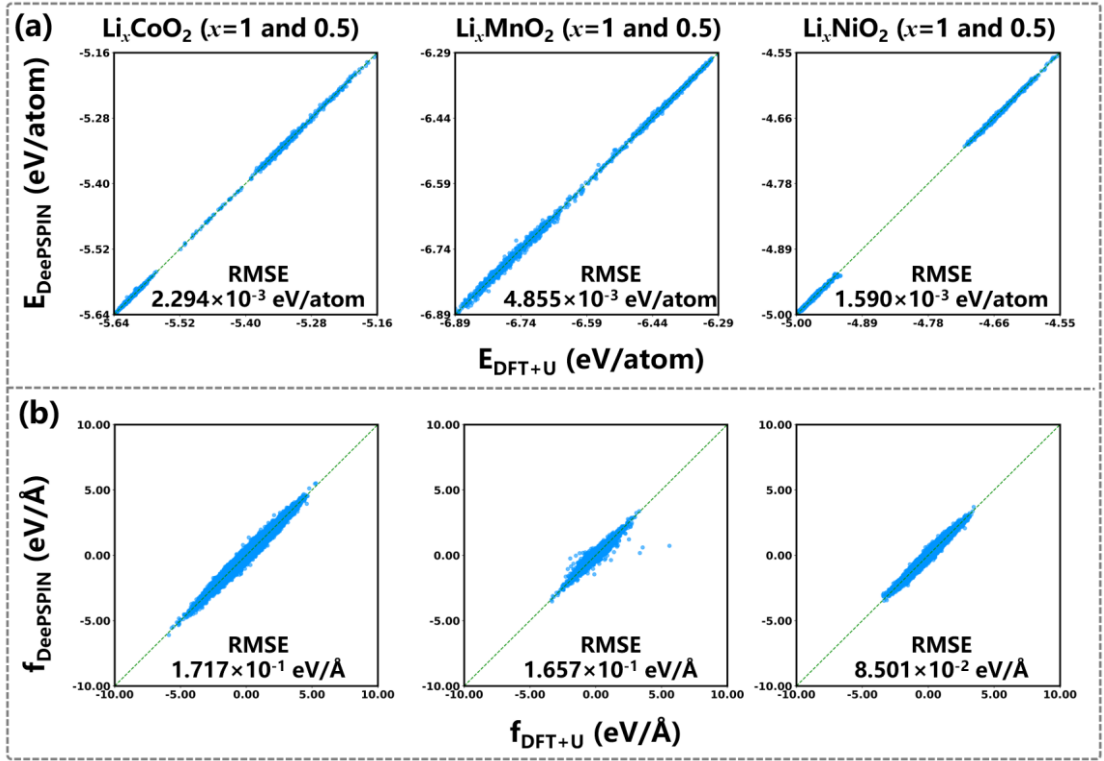

**Figure 4.** Comparisons of the (a) energies and (b) forces predicted by the DFT+U calculations vs. the collinear DeePSPIN model for $Li_xCoO_2$, $Li_xMnO_2$ and $Li_xNiO_2$ ($x = 0.5$ and $1.0$) cathodes. In addition to the spin ground states' data, the spin excited states' data are also included.

We have already performed systematic tests to evaluate the robustness of the collinear DeePSPIN model for the complex TM oxide cathode materials MLPs training. Because of the introduction of spin features into the descriptor, the collinear DeePSPIN model is able to distinguish potential energies associated different spin states. Before applying the collinear DeePSPIN model to more complicated application systems, we need to address one more question: can the collinear DeePSPIN model provide accurate predictions for DFT+U data obtained from completely random initial-guess wavefunctions? Here, we note that a completely random initial wavefunction refers to the situation that neither the initial magnetic moments are set nor the total magnetic moment is controlled in DFT+U single-point calculations. Taking the $LiCoO_2$ and $LiNiO_2$ as examples, we compare the performance of the pristine DP and the collinear DeePSPIN models (see **Figure S2**). We can see that the pristine DP model provides rather poor predictions, especially for the $LiCoO_2$. The RMSEs for energies and forces are ~ 11 meV/atom and 229 meV/Å, respectively. In contrast, the collinear DeePSPIN model provides a much better description for both energies and forces (the RMSEs of energies and forces are ~ 2.8 meV/atom and 130 meV/Å for the $Li_xCoO_2$, and ~ 2.0 meV/atom and 87 meV/Å for the $Li_xNiO_2$, respectively). More importantly, these tests also demonstrate that our proposed collinear DeePSPIN scheme can fully utilize all data generated from DFT+U single-point calculations, regardless of the magnetic states being in ground states or not.

As a final example, we apply the collinear DeePSPIN model to more complicated cathode materials to further verify the model's robustness. Here, we consider the following three cases: Li-Ni anti-site defects in the $LiNiO_2$, the Co-free high-Ni binary $Li_xNi_{1.5}Mn_{0.5}O_4$ ($x$=1.5 and 0.5) and the ternary $Li_xNi_{1/3}Co_{1/3}Mn_{1/3}O_2$ ($x$=1.0 and 0.67) cathode (atomic structures are displayed in **Figure 1c**) materials (please refer to SI for the details of the training datasets construction). We emphasize that the completely random initial wavefunctions were used in all DFT+U calculations. We can see that even for such complex cases, our collinear DeePSPIN model still well reproduces the energies and forces given by DFT+U calculations (see **Figure 5**). While the pristine DP model exhibits lower accuracy, especially for the energies (see **Figure S3**). We note that the collinear DeePSPIN model's accuracy can be further improved by using DPGEN[44] concurrent learning process to actively collect the dataset. Therefore, the collinear DeePSPIN model performs as an effective and accurate tool to construct MLPs for complex cathode materials.

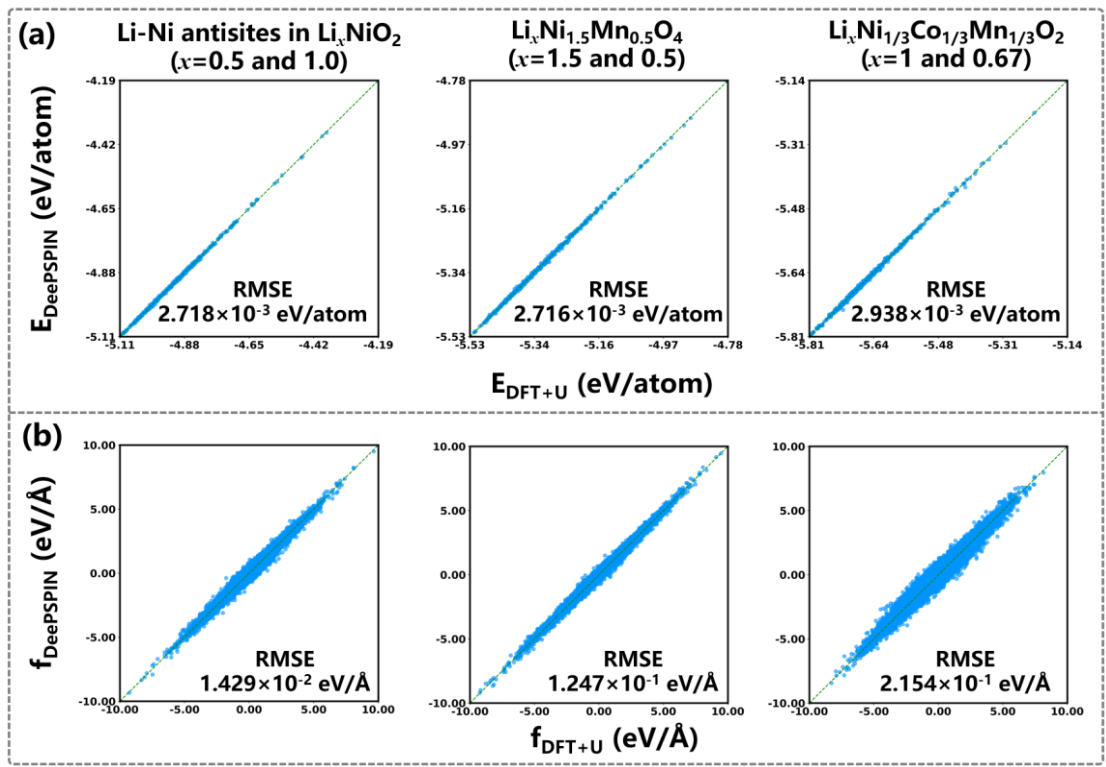

**Figure 5.** Comparisons of the (a) energies and (b) forces predicted by the DFT+U calculations vs. the collinear DeePSPIN model for Li-Ni anti-site defects in LiNiO$_2$, Li$_x$Ni$_{1.5}$Mn$_{0.5}$O$_4$ ($x$ = 1.5 and 0.5) and Li$_x$Ni$_{1/3}$Co$_{1/3}$Mn$_{1/3}$O$_2$ ($x$ = 1 and 0.67).

## Conclusion and Outlook

In this work, we develop a new deep neural network framework based on the pristine DP model by incorporating the atomic spin feature into the descriptor to distinguish different spin states of TM ions in complex cathode materials. We employ a series of test systems, from simple LiTMO$_2$ (TM=Ni, Co, Mn) to complex ternary NCM cathode materials, to justify the accuracy of our proposed collinear DeePSPIN model, which is demonstrated to be able to well reproduce energies and forces obtained by DFT+U calculations. More importantly, all results generated by DFT+U single-point calculations can be utilized and included in the dataset for training MLPs, regardless of the spin configurations being in a ground state or not. Overall, our proposed scheme provides a promising tool to efficiently train MLPs for complex TM oxide cathode materials.

Upon obtaining a robust MLP, we then need to perform MD simulations based on this MLP to derive a dynamic trajectory. In the current force-field model, the potential energies of the spin ground and excited states are all included and the spin degree of freedom serves as an independent variable. We therefore need to conduct energy minimization within the spin subspace at each specific atomic structure to achieve the ground states' energies and forces information along a MD simulation pass. Since the spin value obtained from collinear DFT+U calculations is a discrete scalar, traditional optimization algorithms for continuous variables may not be suitable. To address this issue, we plan to try two possible strategies. The first way is that we can use the automatic differentiation technology of the neural network to calculate the derivative of the energy with respect to the spin degree of freedom to yield "forces" on spin states of TM ions. Owing to lack of spin forces' labels in the training dataset, the values of spin forces obtained in this way may not be sufficiently accurate. However, they can still provide some guidance for our optimization directions of changing spin states. The second strategy is to use global optimization algorithms, such as the genetic algorithm, the particle swarm algorithm, etc., to minimize the energy with respect to the spin degree of freedom considering its discrete characteristic. Once the energy minimization with respect to the spin degree of freedom is completed, we can perform the conventional MD simulations and interface the collinear DeePSPIN model to the DPGEN concurrent learning framework to actively collect dataset. We are working on the development of these approaches.

## Acknowledgements

The authors gratefully acknowledge funding support from the DP Technology Corporation (Grant No. 2021110016001141), Chinese Ministry of Science and Technology (Grant No. 2021YFB3800303), the National Natural Science Foundation of China (Grant No. 52273223), the School of Materials Science and Engineering at Peking University, and the AI for Science Institute, Beijing (AISI). The computing resource of this work was provided by the Bohrium Cloud Platform (https://bohrium.dp.tech), which is supported by DP Technology.

# Supporting Information

# A Spin-dependent Machine Learning Framework for Transition Metal Oxide Battery Cathode Materials

Taiping Hu[1,2], Teng Yang[3], Jianchuan Liu[4], Bin Deng[5], Zhengtao Huang[3], Xiaoxu Wang[5], Fuzhi Dai[2], Guobing Zhou[6], Fangjia Fu[2,7], Ping Tuo[2], Ben Xu[*,3], and Shenzhen Xu[*,1,2]

[1]Beijing Key Laboratory of Theory and Technology for Advanced Battery Materials, School of Materials Science and Engineering, Peking University, Beijing 100871, People's Republic of China

[2]AI for Science Institute, Beijing 100084, People's Republic of China

[3]Graduate School of China Academy of Engineering Physics, Beijing 100088, People's Republic of China

[4]HEDPS, CAPT, College of Engineering and School of Physics, Peking University, Beijing 100871, People's Republic of China

[5]DP Technology, Beijing 100080, People's Republic of China

[6] Institute of Advanced Materials, Jiangxi Normal University, Nanchang 330022, People's Republic of China

[7] School of Mathematical Sciences, Peking University, Beijing 100871, People's Republic of China

Corresponding author: bxu@gscaep.ac.cn, xushenzhen@pku.edu.cn

## Computational Details

### Generation of the Dataset

The number of frames in the training dataset for each system is listed in Table S1 and S2.

**Li$_x$CoO$_2$/Li$_x$MnO$_2$/Li$_x$NiO$_2$ ($x$ = 0.5 and 1.0)**

The datasets for the LiCoO$_2$ and Li$_{0.5}$CoO$_2$ were extracted from our previous work.[1] For the Li$_x$MnO$_2$ and Li$_x$NiO$_2$, we followed the workflow similar with our previous work to obtained the datasets. The low spin states of the Co$^{3+}$, Co$^{4+}$ Ni$^{3+}$ and Ni$^{4+}$ are energetically favorable, while the high spin states of the Mn$^{3+}$ and Mn$^{4+}$ are more stable.[2-4] We used the 2×2×1 supercell for all systems (chemical formula is Li$_x$TM$_{12}$O$_{24}$). We employed the method similar to our previous work to obtain the dataset for the spin ground state and excited states. Specifically, we set the initial magnetic moments guess and control the total magnetic moment by using the MAGMOM and NUPDOWN keywords in the VASP input file.

**Li-Ni anti-defects in LiNiO$_2$**

Based on the optimized structure of the LiNiO$_2$, we carried out random exchanges for Li-Ni pairs. Subsequently, the atomic positions and unit cell were fully relaxed. The forces of each direction for every atom in the modeling supercells were converged within 0.01 eV/Å. We then added random perturbations for the atomic positions and the unit cell to generate about 200 configurations. Finally, self-consistent field calculations were performed for those structures to obtain the energy, force and projected magnetic moments. We note that the random initial wavefunctions were used in all DFT+U single-point calculations.

**Li$_x$Ni$_{1.5}$Mn$_{0.5}$O$_4$ ($x$ = 0.5 and 1.5) and Li$_x$Ni$_{1/3}$Co$_{1/3}$Mn$_{1/3}$O$_2$ ($x$ = 1 and 0.67)**

We obtained the initial structures of the LiNi$_{1.5}$Mn$_{0.5}$O$_4$ from the materials project[5] and the LiNi$_{1/3}$Co$_{1/3}$Mn$_{1/3}$O$_2$ from the previous work.[6, 7] For the Li$_x$Ni$_{1/3}$Co$_{1/3}$Mn$_{1/3}$O$_2$ case, we used a supercell with chemical formular of $Li_{27}Ni_9Co_9Mn_9O_{54}$, therefore a 2×2×1 Monkhorst−Pack k-point mesh was used to sample the Brillouin zone. Both of the lattice parameters and the ionic positions were fully relaxed during the optimizations. We then created several Li-vacancy structures by randomly removing a specific number of Li ions. All those Li-vacancy structures were further fully relaxed. The forces of each direction for every atom in the modeling supercells were converged within 0.01 eV/Å in the structural relaxation calculations. We then added perturbations to the ionic position and the unit cell to generated configurations for each system. Finally, self-consistent field calculations were performed for those structures to obtain energies, forces and projected magnetic moments. We note that the random initial wavefunctions were used in all DFT+U calculations.

# Figures

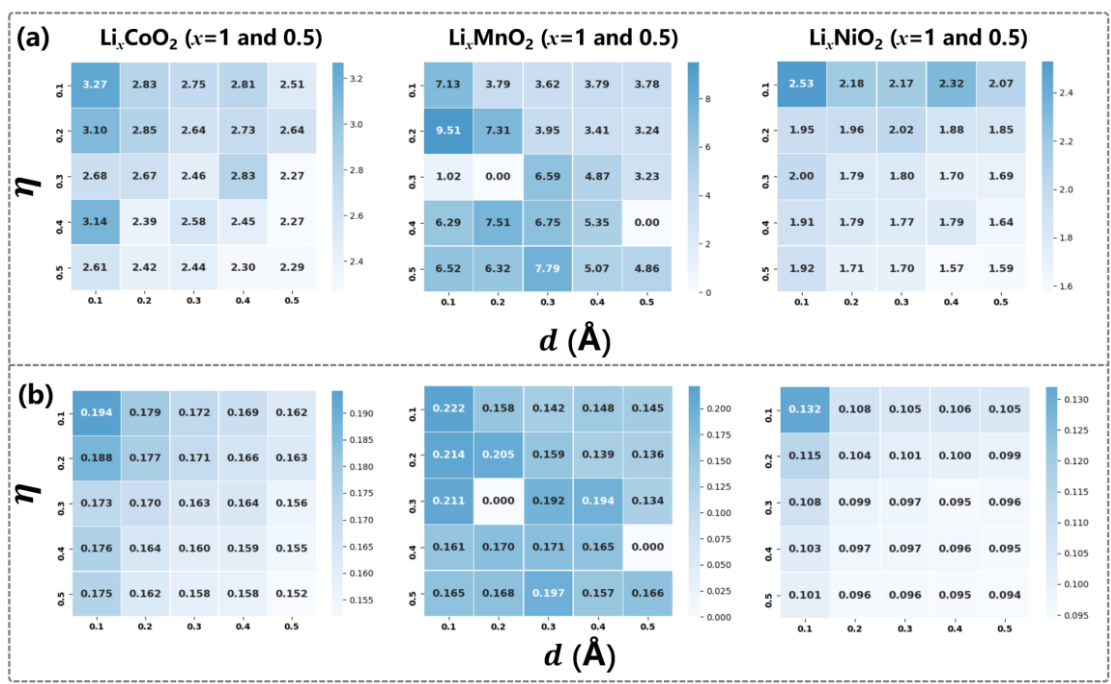

**Figure S1.** RMSEs of the (a) energies (meV/atom) and (b) forces (eV/Å) predicted by DFT+U calculations vs. the collinear DeePSPIN models with different $\eta$ and $d$ parameters for $Li_xCoO_2$, $Li_xMnO_2$ and $Li_xNiO_2$ cathodes. The numerical value 0 indicates that the deepmd-kit software reports atomic overlaps during dataset's checking.

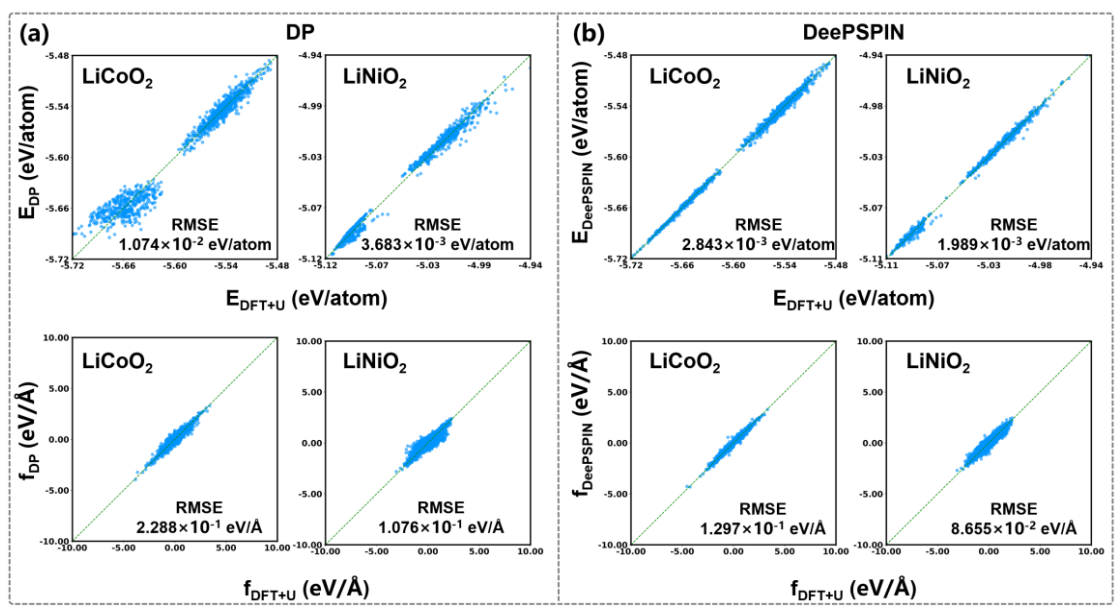

**Figure S2.** Comparisons of the energies (the up row) and forces (the bottom row) predicted by the DFT+U calculations vs. (a) the pristine DP models and vs. (b) the collinear DeePSPIN

S5

models for LiCoO$_2$ and LiNiO$_2$ cases. We used random initial wavefunctions in all DFT+U single-point calculations.

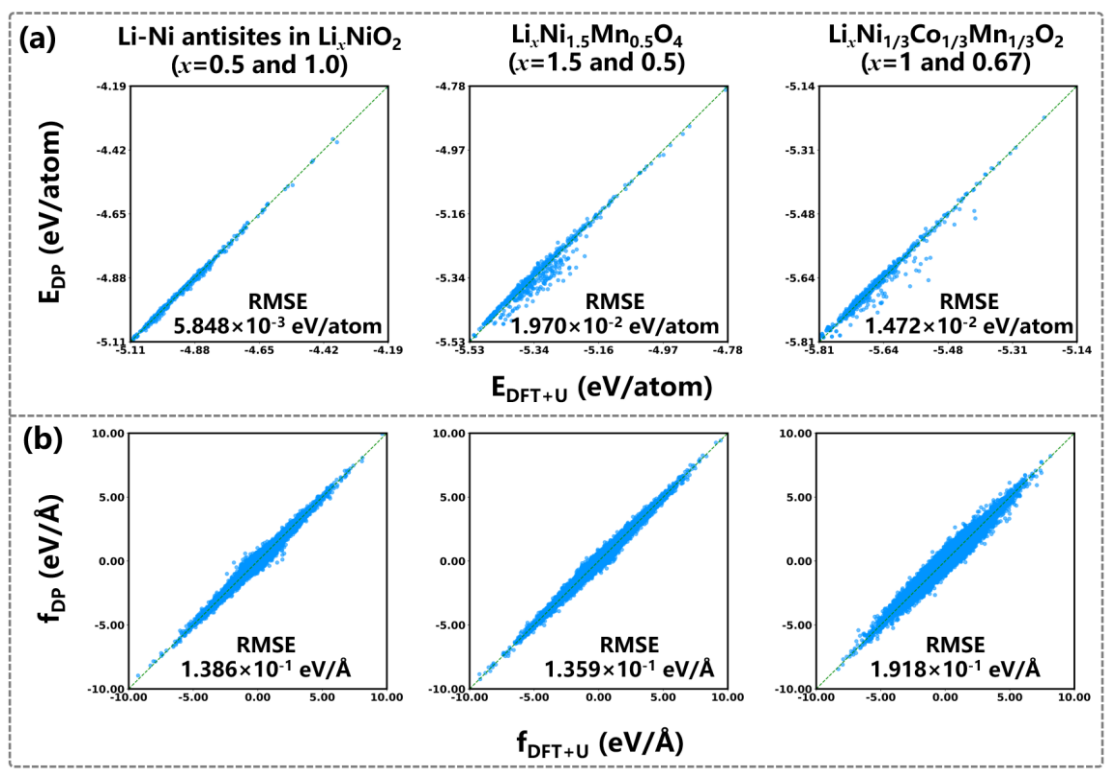

**Figure S3.** Comparisons of the (a) energies and (b) forces predicted by the DFT+U calculations vs. the pristine DP model for Li-Ni anti-site defects in LiNiO$_2$, Li$_x$Ni$_{1.5}$Mn$_{0.5}$O$_4$ ($x$=1.5 and 0.5) and Li$_x$Ni$_{1/3}$Co$_{1/3}$Mn$_{1/3}$O$_2$ ($x$=1 and 0.67). We used random initial wavefunctions in all DFT+U single-point calculations.

# Tables

**Table S1.** The numbers of frames used for MLPs training in the Li$_x$TMO$_2$. The left side and right side of the plus sign represent $x=0.5$ and $x=1.0$, respectively.

|  | Li$_x$CoO$_2$ ($x$=0.5 and 1.0) | Li$_x$MnO$_2$ ($x$=0.5 and 1.0) | Li$_x$NiO$_2$ ($x$=0.5 and 1.0) |
|---|---|---|---|
| Spin ground state | 675+438 | 960+996 | 955+979 |
| Spin excited state | 630+437 | 873+987 | 929+930 |

**Table S2.** The numbers of frames used for MLPs training in the Li-Ni anti-sites in LiNiO$_2$, Li$_x$Ni$_{1.5}$Mn$_{0.5}$O$_4$, and Li$_x$Ni$_{1/3}$Co$_{1/3}$Mn$_{1/3}$O$_2$. The meaning of the plus sign is the same as above (Table S1).

| Li-Ni anti-sites | Li$_x$Ni$_{1.5}$Mn$_{0.5}$O$_4$ ($x$=0.5 and 1.5) | Li$_x$Ni$_{1/3}$Co$_{1/3}$Mn$_{1/3}$O$_2$ ($x$=0.67 and 1.0) |
|---|---|---|
| 198 | 479+156 | 436+186 |

**Table S3.** Tests of the end-pre-factor of regular atomic forces in the loss function. The start-pre-factors of forces in all cases were set as 1000.

|  |  | 1.0 | 5.0 | 10.0 | 100.0 | 500.0 |
|---|---|---|---|---|---|---|
| Li$_x$CoO$_2$ | Energy | $2.150\times10^{-3}$ | $2.233\times10^{-3}$ | $2.294\times10^{-3}$ | $2.679\times10^{-3}$ | $3.633\times10^{-3}$ |
|  | Force | $1.56\times10^{-1}$ | $1.539\times10^{-1}$ | $1.717\times10^{-1}$ | $1.530\times10^{-1}$ | $1.393\times10^{-1}$ |
| Li$_x$MnO$_2$ | Energy | $6.034\times10^{-3}$ | $5.987\times10^{-3}$ | $4.855\times10^{-3}$ | $6.233\times10^{-3}$ | $9.328\times10^{-3}$ |
|  | Force | $2.150\times10^{-1}$ | $1.984\times10^{-1}$ | $1.657\times10^{-1}$ | $1.943\times10^{-1}$ | $1.948\times10^{-1}$ |
| Li$_x$NiO$_2$ | Energy | $1.530\times10^{-3}$ | $1.685\times10^{-3}$ | $1.590\times10^{-3}$ | $1.904\times10^{-3}$ | $2.065\times10^{-3}$ |
|  | Force | $9.739\times10^{-2}$ | $9.554\times10^{-2}$ | $8.501\times10^{-2}$ | $9.138\times10^{-2}$ | $8.572\times10^{-2}$ |